\documentclass[longauth]{aa}  

\usepackage{graphicx}
\usepackage{txfonts}
\usepackage{color}
\begin{document}

   \title{X-ray polarization measurement of the gold standard of radio-quiet active galactic nuclei : NGC~1068}

   \titlerunning {X-ray polarimetry of NGC~1068}

  \author{
F. Marin \inst{1}\thanks{E-mail: frederic.marin@astro.unistra.fr}
\and          
A. Marinucci \inst{2}
\and          
M. Laurenti \inst{3,4,17}
\and          
D. E. Kim \inst{5,6,3}
\and          
T. Barnouin \inst{1}
\and          
A. Di Marco \inst{5}
\and          
F. Ursini \inst{7}
\and          
S. Bianchi \inst{7}
\and          
S. Ravi \inst{8}
\and          
H.L. Marshall \inst{8}
% ...
}

   \authorrunning{F. Marin et al.}

  \institute{Université de Strasbourg, CNRS, Observatoire Astronomique de Strasbourg, UMR 7550, 67000 Strasbourg, France 
         \and
         ASI - Agenzia Spaziale Italiana, Via del Politecnico snc, 00133 Roma, Italy 
         \and
         Dipartimento di Fisica, Università di Roma “Tor Vergata”, Via della
Ricerca Scientifica 1, 00133 Roma, Italy 
         \and
         Space Science Data Center, SSDC, ASI, Via del Politecnico snc,
00133 Roma, Italy 
         \and
         INAF Istituto di Astrofisica e Planetologia Spaziali, Via del Fosso del Cavaliere 100, I-00133 Roma, Italy 
         \and
         Dipartimento di Fisica, Università degli Studi di Roma “La Sapienza”, Piazzale Aldo Moro 5, I-00185 Roma, Italy 
         \and
         Dipartimento di Matematica e Fisica, Università degli Studi Roma
Tre, via della Vasca Navale 84, 00146 Roma, Italy 
         \and
         MIT Kavli Institute for Astrophysics and Space Research, Massachusetts Institute of Technology, 77 Massachusetts Avenue, Cambridge,
MA 02139, USA 
         \and
         Science and Technology Institute, Universities Space Research Association, Huntsville, AL 35805, USA 
         \and
         Université Grenoble Alpes, CNRS, IPAG, 38000 Grenoble, France 
         \and
         School of Mathematics, Statistics, and Physics, Newcastle University, Newcastle upon Tyne NE1 7RU, UK 
         \and
         Center for Astrophysics, Harvard \& Smithsonian, 60 Garden St., Cambridge, MA 02138, USA 
         \and
         Astronomical Institute of the Czech Academy of Sciences, Boční II
1401/1, 14100 Praha 4, Czech Republic 
         \and
         Istituto Nazionale di Fisica Nucleare, Sezione di Roma "Tor
Vergata", Via della Ricerca Scientifica 1, 00133 Roma, Italy 
         \and
         Department of Astronomy, University of Maryland, College Park,
Maryland 20742, USA 
         \and
         Instituto de Astrofísica de Andalucía—CSIC, Glorieta de la
Astronomía s/n, 18008 Granada, Spain 
         \and
         INAF Osservatorio Astronomico di Roma, Via Frascati 33, 00078
Monte Porzio Catone (RM), Italy 
% ...
}
    
   \date{Received February X, 2024; accepted Month Day, 2024}

 
  \abstract
   {NGC~1068 is the radio-quiet active galactic nucleus (AGN) which has been the most observed in polarimetry. Yet, its high energy polarization has never been probed before due to the lack of dedicated polarimeters.}
   {Using the first X-ray polarimeter sensitive enough to measure the polarization of AGNs, we want to probe the orientation and geometric arrangement of (sub)parsec-scale matter around the X-ray source.}
   {We used the Imaging X-ray Polarimetry Explorer ({\it IXPE}) satellite to measure, for the first time, the 2-8 keV polarization of NGC~1068. We pointed {\it IXPE} for a net exposure time of 1.15~Ms on the target, in addition to two $\sim$ 10~ks each {\it Chandra} snapshots in order to account for the potential impact of several ultraluminous X-ray source (ULXs) within {\it IXPE}'s field-of-view.}
   {We measured a 2 -- 8~keV polarization degree of 12.4\% $\pm$ 3.6\% and an electric vector polarization angle of 101$^\circ$ $\pm$ 8$^\circ$ at 68\% confidence level. If we exclude the spectral region containing the bright Fe~K lines and other soft X-ray lines where depolarization occurs, the polarization fraction rises up to 21.3\% $\pm$ 6.7\% in the 3.5 -- 6.0~keV band, with a similar polarization angle. The observed polarization angle is found to be perpendicular to the parsec scale radio jet. Using a combined {\it Chandra} and {\it IXPE} analysis plus multi-wavelength constraints, we estimated that the circumnuclear "torus" may sustain a half-opening angle of 50 -- 55$^\circ$ (from the vertical axis of the system).}
   {Thanks to {\it IXPE}, we have measured the X-ray polarization of NGC~1068 and found comparable results,  both in terms of polarization angle orientation with respect to the radio-jet and torus half-opening angle, to the X-ray polarimetric measurement achieved for the other archetypal Compton-thick AGN : the Circinus galaxy. Probing the geometric arrangement of parsec-scale matter in extragalactic object is now feasible thanks to X-ray polarimetry.}


   \maketitle
%

\section{Introduction}
\label{Introduction}

NGC~1068, also known as Messier~77 (M77), is an optically bright (V = 11.8~mag in a 4.9$''$ aperture, \citealt{Sandage1973}) type-2 active galactic nuclei (AGN), located in the nearby Universe ($z$ = 0.00379). Because of its proximity and brightness, it is among the AGNs that have been the most studied over the past century. In particular, it is the source that is at the origin of the Unified Scheme of AGNs such as established by \citet{Antonucci1993} and \citet{Urry1995}. In this zeroth-order model, all AGNs are similar in morphology, but their apparent observational properties can differ depending on the inclination of the AGN core with respect to the observer. This is mainly due to the fact that, along the equatorial plane of the AGN, lies an optically-thick circumnuclear reservoir of gas and dust that creates a strong geometric asymmetry in emission and absorption. The model has evolved since then, encompassing evolution and accretion to explain several observational differences that orientation cannot explicate alone \citep{Dopita1997}, but determining the real inclination and morphology of AGNs remains a challenge.

In the case of NGC~1068, there seems to be a consensus about its nucleus inclination. Studies based on the modelling of the spectral energy distribution usually find a (torus) inclination of the order of 70 -- 90$^\circ$ \citep{Honig2007,Honig2008,Lopez2018}. Observations and modeling of the polar outflows kinematics bring more extreme inclination values, $\sim$ 85$^\circ$, under the assumption of biconical outflow models \citep{Das2006,Fischer2013,Miyauchi2020}. Interestingly, Fig.~8b in \citet{Miyauchi2020} shows the 0.1$''$ ($\sim$ 7~pc) scale distribution of the gas clouds in NGC~1068 that has been essentially confirmed by the VLTI/MATISSE reconstructed image obtained by \citet{Gamez2022}, who also estimate that the circumnuclear, optically thick, dusty region is seen nearly edge-on. Finally, the presence of water masers coexisting with the central engine in NGC~1068 can be used to trace the structure and velocity field of the torus, provided that the dusty molecular torus is indeed responsible for the maser emission \citep {Greenhill1996}. VLA observations and Effelsberg 100m monitoring allowed to estimate that the maser disk is rather thin and inclined by 80 -- 90$^\circ$ \citep{Gallimore2001}. VLBI observations of the same water maser also allowed to estimate the central black hole mass ($\sim$ 10$^7$~M$_\odot$, \citealt{Greenhill1996}).  

Polarization, being responsive to the morphology of a source, serves as an independent tool for deducing the orientation of AGNs. Using three dimensional models of the system, \citet{Packham1997} and \citet{Kishimoto1999} attempted to reproduce their near-infrared and near-ultraviolet polarization maps of NGC~1068, respectively. Both found that the northern outflowing region is inclined towards the observer, while the southern wind is directed away, but could not determine the AGN inclination nor the torus geometry. The main problem with this approach is that the ultraviolet, optical and infrared bands are heavily polluted by unpolarized starlight emission from the host, dust reemission by the torus and vigorous starburst activity \citep{Romeo2016}. To get around this problem, it is necessary to go to higher energies, where there is neither stellar emission nor re-emission by hot dust. Thanks to the very low contribution of polluting X-ray sources, and therefore to a rather clean observational window, X-ray polarimetry offers a powerful probe to narrow down the uncertainties in determining the AGN inclination, together with its torus half-opening angle and density \citep{Goosmann2011,Marin2018c}. 

It is now possible to measure the X-ray polarization of extragalactic objects thanks to the Imaging X-ray Polarimetry Explorer ({\it IXPE}), NASA’s first mission to study the polarization of astronomical X-rays that was launched in December 2021 \citep{Weisskopf2022}. {\it IXPE} is capable of measuring the linear polarization of electromagnetic waves from 2 to 8~keV, and it was already pointed towards several radio-quiet AGNs : NGC~4151 \citep{Gianolli2023}, MCG-05-23-16 \citep{Marinucci2022,Tagliacozzo2023}, IC~4329A \citep{Ingram2023,Pal2023} and the Circinus galaxy \citep{Ursini2023}. The later observation revealed that the neutral reflector in Circinus produces a polarization degree of 28\% $\pm$ 7\% at 68\% confidence level, with a polarization angle roughly perpendicular to the radio jet. According to Monte Carlo simulations, the half-opening angle of the torus was determined to be of the order of 50$^\circ$ $\pm$ 5$^\circ$ \citep{Ursini2023}. 

Indeed, polarimetry has the ability to constrain the structure of the various regions surrounding the continuum  source as well as the orientation of the object with respect to the observer \citep{Tinbergen2005}. X-ray polarimetry, in particular, can probe the geometry of the molecular torus with high precision, as shown in \citet{Marin2016}, \citet{Marin2018} or \citet{Podgorny2024}. The reason for this is that the emitting, scattering and absorbing signatures of the torus can be clearly identified by X-ray spectroscopy thanks to spectral decomposition, something that is not easily achievable at longer wavelengths. Thus, if we measure the polarization of the torus in X-rays, it becomes possible to have strong geometric and orientation constraints on the overall morphology of this compact region by comparing the data to simulations, as shown in the Circinus case \citep{Ursini2023}.

In the hopes to achieve similar results, we hereby report the first pointing of NGC~1068 by {\it IXPE}, together with two {\it Chandra} snapshots, to enable a spatially resolved spectro-polarimetric study. For the remainder of this paper, a $\Lambda$CDM cosmology with $H_0 = 70$ km s$^{-1}$ Mpc$^{-1}$, $\Omega_\mathrm{m}=0.3$ and $\Omega_\Lambda = 0.7$ is adopted. Also, unless specified otherwise, errors are given at 68\% confidence level for one parameter of interest and upper limits are given at 99\% confidence level.

\section{Observations and data reduction}
\label{Observation}

\subsection{{\it IXPE}}
\label{Observation:IXPE}
{\it IXPE} observed NGC~1068 (02h42m40.711s, -00d00m47.81s in J2000.0 equatorial coordinates) from January 3 to January 29, 2024 (ObsID 02008001), for a net exposure time of 1.15~Ms. Cleaned level 2 event files were firstly treated with the background rejection procedure described in \cite{DiMarco2023}\footnote{The adopted script is available on https://github.com/aledimarco/IXPE-background}. Then they were produced and calibrated using standard filtering criteria with the dedicated {\sc FTOOLS} tasks and the latest calibration files available in the {\it IXPE} calibration data base (CALDB 20230526). 

I, Q, and U spectra were extracted according to the weighted analysis method described in \cite{DiMarco2022} by setting the parameter \texttt{stokes=neff} in {\sc XSELECT}. The background spectra were extracted from source-free circular regions with a radius of 78$''$. Extraction radii for the I Stokes spectra of the source were computed via an iterative process which leads to the maximization of the signal-to-noise ratio (SNR) in the 2 -– 8~keV energy band \citep[similarly to][]{Piconcelli2004}, which led to radii of 47$''$, 52$''$ and 47$''$ for the Detector Units (DUs) 1, 2 and 3, respectively. The same circular regions (centered on the source) were used for all three DUs for I, Q and U. This results in a 2-5.5 keV background of 10.4\%, 14.9\% and 13.9\% for the total DU1, DU2, and DU3 I spectra. These values rise to 23.7\%, 28.4\% and 25.7\% in the 5.5 -- 8~keV band. We used a constant energy binning of 0.2~keV for Q and U Stokes spectra and required an SNR higher than 3 in the intensity spectra. The choice of dividing the full energy band into two bins below and above 5.5~keV is driven by the need to explore the influence of the iron line complex that dominates the emission above 5.5~keV.

\subsection{{\it Chandra}}
\label{Observation:Chandra}

Similarly to the Circinus case \citep{Ursini2023}, there are several ultraluminous X-ray (ULX) sources around NGC~1068 \citep{Smith2003, Matt2004, Zaino2020}. To monitor the flux levels of the off-nuclear point like sources, two {\it Chandra} observations were performed at the beginning and at the end of the {\it IXPE} pointings, on 2024, January 4 (ObsID 29071) and on January 28 (ObsID 29072), with the Advanced CCD Imaging Spectrometer \citep[ACIS,][]{Garmire2003}. To reduce pileup effects, the frame time was set to 0.5 s and custom CCD subarrays were used. Data were reduced with the Chandra Interactive Analysis of Observations \citep[CIAO,][]{Fruscione2006} 4.14 and the Chandra Calibration Data Base 4.11.0, adopting standard procedures. We generated event files for the two observations with the {\sc CIAO} tool {\tt chandra$_{-}$repro} and, after cleaning for background flaring, we got net exposure times of 10.3 ks and 9.3 ks for the first and for the second observation, respectively.

We used a circular extraction region with a 15$''$ radius for the AGN and background spectra were extracted from source-free circular regions with a radius of 10$''$. Spectra from the two observations were added and then binned in order to oversample the instrumental resolution by a factor of 3 and to have no less than 30 counts in each background-subtracted spectral channel. This allows the applicability of the $\chi^2$ statistic. We ignored channels between 8 and 10 keV due to pileup, which was calculated via the {\sc CIAO} tool {\tt pileup$_{-}$map}. We estimated an average pileup fraction of 7.5\% in the central $3\times3$ pixels region, ranging from 5\% to 10\% (in the central pixel) for both observations. 

Spectra from other point-like sources within the field of view were extracted from circular regions with 2.5$''$ radii. In the 2-8 keV band, the brightest ones are: SN 2018ivc \citep{Bostroem20, Maeda23}, CXOU J024241.4-000013, CXOU J024238.9-000055 and CXOU J024241.0-000125 \citep{Smith2003}. Spectral fit of the other point-like sources showed 2-8 keV fluxes lower than $\sim2.5\times10^{-14}$ erg cm$^{-2}$ s$^{-1}$, which is approximately 0.5\% of the total flux of the AGN. These sources were not considered in the spectro-polarimetric fit, since their overall contribution is negligible. Spectra were then binned requesting at least 5 counts in each spectral bin and the Cash statistics \citep{Cash79} was used for the data analysis.

\begin{figure}
\centering
\includegraphics[width=\columnwidth]{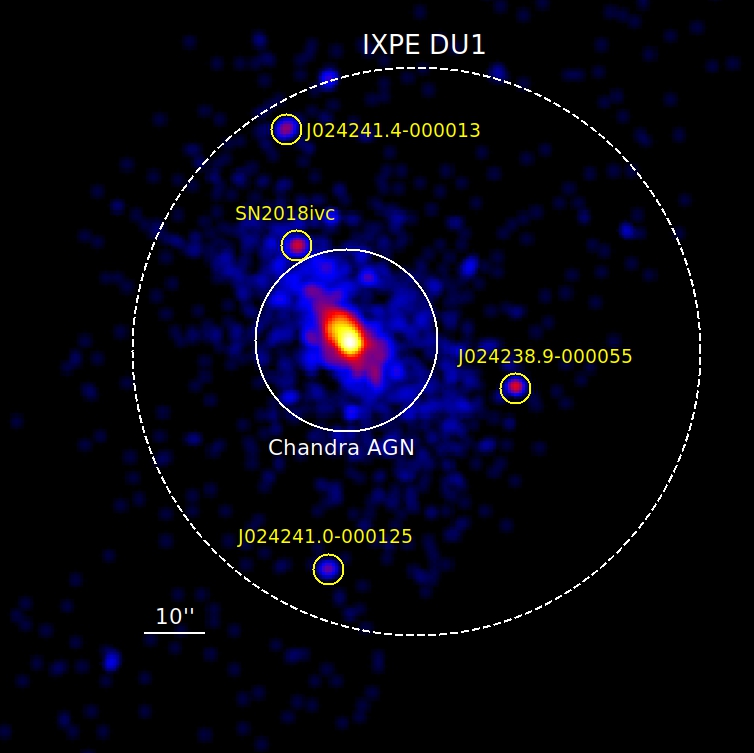}
\caption{The 2.\arcmin $\times$ 2.\arcmin central region of NGC 1068 is shown in the 0.3-7.0 keV band (\textit{Chandra} ObsID 29071). The dashed white circle corresponds to the {\it IXPE} DU1 source extraction region. A Gaussian smoothing kernel of $3\times3$ pixels has been applied for the for the sake of visual clarity.}
\label{fig:chandra-image}
\end{figure}


\section{Measures of NGC~1068's X-ray polarization}
\label{Results}

\subsection{{\it IXPE} polarization cubes}
\label{Results:IXPE_pcubes}

We estimated the X-ray polarization properties using event-based Stokes parameter analysis methods, as implemented in \cite{Kislat2015}. The model-independent analysis was performed using the unweighted method employing the \texttt{PCUBE} algorithm in \texttt{ixpeobssim}, as detailed in \cite{Baldini2022}. The polarization degree and angle were derived from normalized $q$ and $u$ Stokes parameters, calculated as {\small $P$ = $\sqrt{({q})^2+({u})^2}$} and {\small $\Psi$ = $1/2\tan^{-1}({u/q})$} (measured from North to East), with the background polarization subtracted. Consequently, we estimated the time-averaged X-ray polarization properties of the source within the 2 -– 8~keV range as $P$ = 9\% $\pm$ 5\% and $\Psi$ = 113$^\circ$ $\pm$ 16$^\circ$. As justified in Sect.~\ref{Observation:IXPE}, we divided the nominal energy band into two bins and measured $P$ = 14\% $\pm$ 4\% and $\Psi$ = 105$^\circ$ $\pm$ 8$^\circ$ (99.7\% confidence level) for the energy range 2 -- 5.5 keV. The remaining {\it IXPE} band, 5.5 -- 8~keV, only shows an upper limit with $P <$ 20\%.

Furthermore, to explore the variability of polarization over time and energy, we segmented the entire observation into subsets based on time and energy, as described in \citet{Kim2024}. Specifically, we divided the $q$ and $u$ data into identical time spans, depending on the selected number of bins from 2 to 15 bins (2 bins with 1.15~Ms give 575 ks/bin; 15 bins correspond to ~ 77 ks/bin). In all cases, the probability of the hypothesis that the data are constant was greater than 28\%, implying no significant evidence of time variability. Energy-resolved analysis was performed in a similar manner. In this case, we divided the 2–8 keV range into 2, 3, 4, 6, 8, and 12 bins (e.g., 2 bins = 2 -– 5~keV, 5 –- 8~keV). The polarization is consistent with being constant with energy, with a probability greater than 24\% in all cases. Additional, minor details are shown in the Appendix regarding the PCUBE analysis.

\subsection{{\it IXPE} spectropolarimetry}
\label{Results:IXPE_spectro}

The X-ray spectropolarimetric analysis was carried out with \textsc{XSPEC} \texttt{v12.13.1e} \citep{Arnaud1996}. The Galactic absorption along the line of sight amounts to $N_\mathrm{H,Gal} = 2.59\times10^{20}$ cm$^{-2}$ \citep{HI4PI} and is accounted for by using the \textsc{tbabs} model with the ISM abundances from \cite{Wilms2000}. We fit the I, Q, and U spectra simultaneously while including a cross-calibration constant (\textsc{const}) between the three different DUs. For what concerns the sole IXPE data, we focus on a phenomenological modeling while a more physical approach is described in the following section. In this case, the X-ray continuum is modeled as a simple power law (\textsc{powerlaw}) throughout the whole feasible 2 -- 8~keV energy interval. 

In order to probe the polarization properties of NGC~1068 we  adopted the \textsc{polconst} model, that assumes an energy-independent polarization and has two free parameters: $P$ and $\Psi$. Thus, we first start with a model that can be written as  $\textsc{const} \times \textsc{tbabs} \times \textsc{polconst} \times \textsc{powerlaw}$ in \texttt{XSPEC}. The spectral fit is clearly unsatisfactory ($\chi^2 / \mathrm{dof}= 931/476$), since we observe large positive residuals at energies larger than 6~keV, likely owing to the broad iron line complex. We thus add a Gaussian emission line (\textsc{zgauss}) to account for this component, and this leads to a significant fit improvement ($\chi^2 / \mathrm{dof}= 537/473$), implying $\Delta{\chi^2}=394$ for three degrees of freedom. We fixed the line width to the value of $\sigma$ = 500 eV returned by the best-fit, since the centroid energy cannot be constrained otherwise. Such a broad emission feature is the result of the blend of several emission lines from both neutral and ionized gas \citep{Matt2004}, that we cannot individually resolve with IXPE. The emission line is centered at $E$ = 6.58$_{-0.04}^{+0.05}$ keV. We find a rather soft X-ray continuum, with the photon index having a value of $\Gamma$ = 2.35$_{-0.04}^{+0.05}$. 

Nonetheless, we still observe some residuals at softer energies which we try to account for by including an additional power-law component. This leads to a further improvement of the spectral fit ($\chi^2 / \mathrm{dof}= 474/472$). Thus, our best-fit phenomenological model consists of: 
\begin{equation*}
    \textsc{const} \times \textsc{tbabs} \times \textsc{polconst} \times ( \textsc{powerlaw}_1 + \textsc{powerlaw}_2 + \textsc{zgauss})
\end{equation*}

\noindent The line centroid shifts towards $E$ = 6.67 $\pm$ 0.06 keV and the two power-law photon indices are $\Gamma_1$ = $0.9 \pm 0.4$ and $\Gamma_2$ = 4.3$_{-0.6}^{+0.8}$. The best-fit values of the 2 -- 8~keV polarization degree and angle are $P$ = 12.4\% $\pm$ 3.6\% and $\Psi$ = 100.7$^\circ$ $\pm$ 8.3$^\circ$, respectively.

\begin{figure}[t]
\centering
\includegraphics[width=\columnwidth]{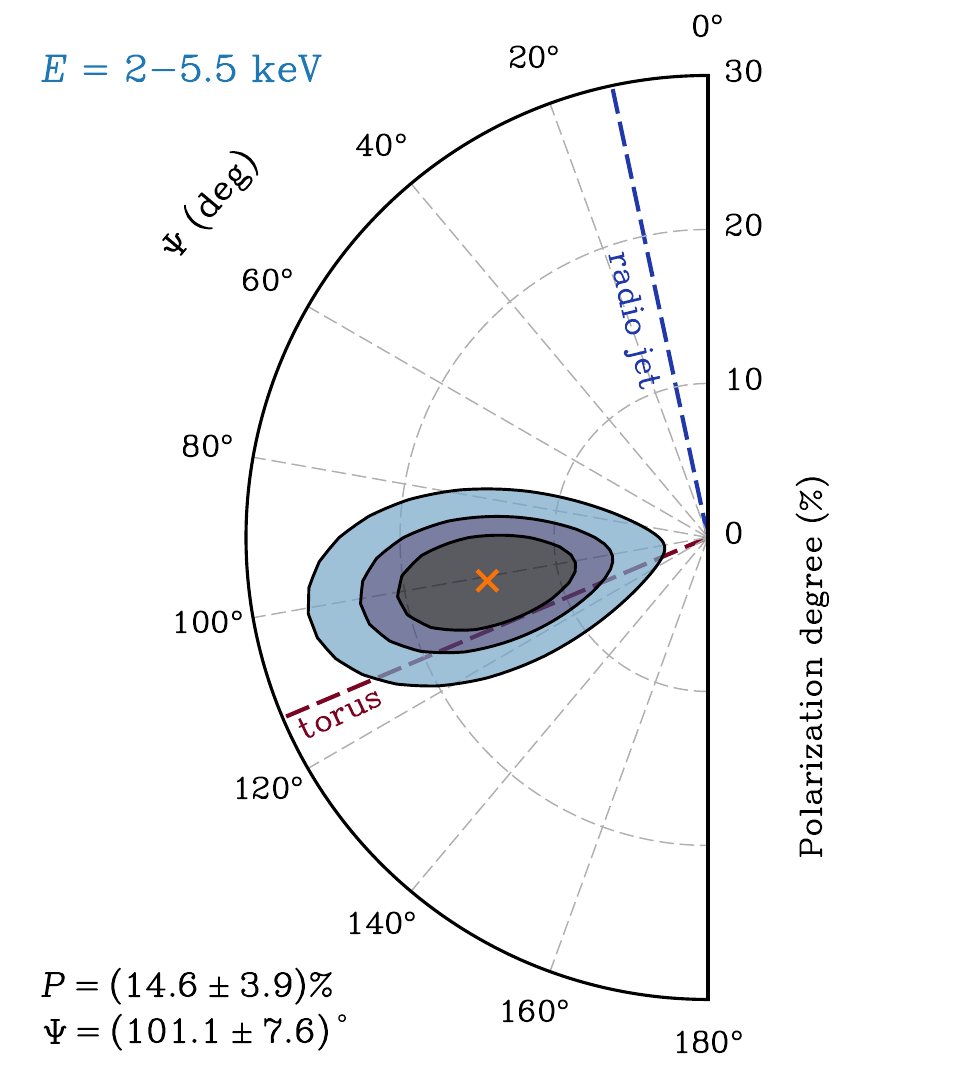}
\caption{Contour plot of the polarization angle vs polarization degree accounting for the 68\%, 90\% and 99\% confidence levels, computed by taking into account the IXPE I, Q and U spectra in the 2 -- 5.5~keV band (the energy band with the highest significance in terms of polarization detection). The orange cross indicates the best-fit values of $P$ and $\Psi$, that are also reported in the lower left corner. The dashed blue line describes the direction of the parsec-scale radio jet \citep{Wilson1983}. The dashed brown line indicates the orientation of the molecular torus derived by \citet{Garcia2019} with {\it ALMA}, see Sect.~\ref{exploitation:Geometry}.}
\label{fig:contours}
\end{figure}

Finally, we divided the broad energy band into smaller intervals to probe the polarization properties of NGC~1068. The 2 -- 5.5~keV band shows the highest significance in terms of polarization detection. Fig.~\ref{fig:contours} shows the contour plot of the polarization angle versus the polarization degree in such interval. At energies higher than 5.5~keV, 
there is a clear rise of background contribution in addition to the emission lines from the iron K complex that are blended and widened due to the modest {\it IXPE} spectral resolution. The results are listed in Tab.~\ref{tab:xspec_intvls} and are generally in good agreement with the values reported from the \texttt{PCUBE} model-independent analysis described in the previous section.

\subsection{{\it IXPE}+{\it Chandra} fitting}
\label{Results:IXPE+Chandra}

Thanks to the {\it Chandra} data, we are able to take into account the contribution of the ULXs in the spectral analysis \citep{Bauer2015}. We co-add the {\it Chandra} spectra of extranuclear point sources that are significantly detected above 2 keV, and we model this combined ULX emission with a power law having a fixed slope of 1. We obtain a 2--8 keV flux of $2.2\times10^{-13}$ erg~cm$^{-2}$~s$^{-1}$ for the ULX power law. The AGN, on the other hand, has a 2 -- 8 keV flux of $4.3\times10^{-12}$ erg~cm$^{-2}$~s$^{-1}$.

The 2--8 keV continuum is well described by two components: a warm reflector, mostly contributing below 4 keV, and a cold reflector, dominating above 5 keV \citep[see also][]{Matt2004}. In both cases, the reflected continuum is expected to be significantly polarized, while the associated emission lines are expected to be unpolarized. We fit jointly the \textit{Chandra} and \textit{IXPE} $I$, $Q$, $U$ Stokes spectra with a phenomenological model consisting of: a power law to describe the warm reflector continuum, a {\sc pexrav} \citep{Magdziarz1995} component for the cold reflector continuum, and six Gaussian emission lines\footnote{We use {\sc pexrav} plus emission lines instead of models which self-consistently include both the continuum and the lines because the polarization of the reflected continuum and of the emission lines is expected to be different, see also \cite{Ursini2023}.}.
The photon indices of the warm and cold reflection components are not well constrained by the {\it Chandra + IXPE} fit, therefore we adopt values consistent with those reported by \cite{Matt2004}, \cite{Marinucci2016} and \cite{Zaino2020}. We fix the photon index of \textsc{pexrav} at 2, and that of the warm reflector at 3\footnote{The soft spectrum is steeper than the primary continuum because it is dominated by recombination and resonant lines \citep{Guainazzi2007}, mostly unresolved in our case. However, we checked that we obtain consistent polarimetric results if we fix the photon index of the warm reflector at 2, or if we leave it free to vary.}. In \textsc{pexrav}, the high-energy cut-off is not included, being fixed at the maximum value of $10^6$ keV because the fit is not sensitive to this parameter. The reflection fraction is fixed at $-1$, which yields the reflection component alone in \textsc{pexrav}.

\begin{figure}[]
\centering
\includegraphics[width=\columnwidth]{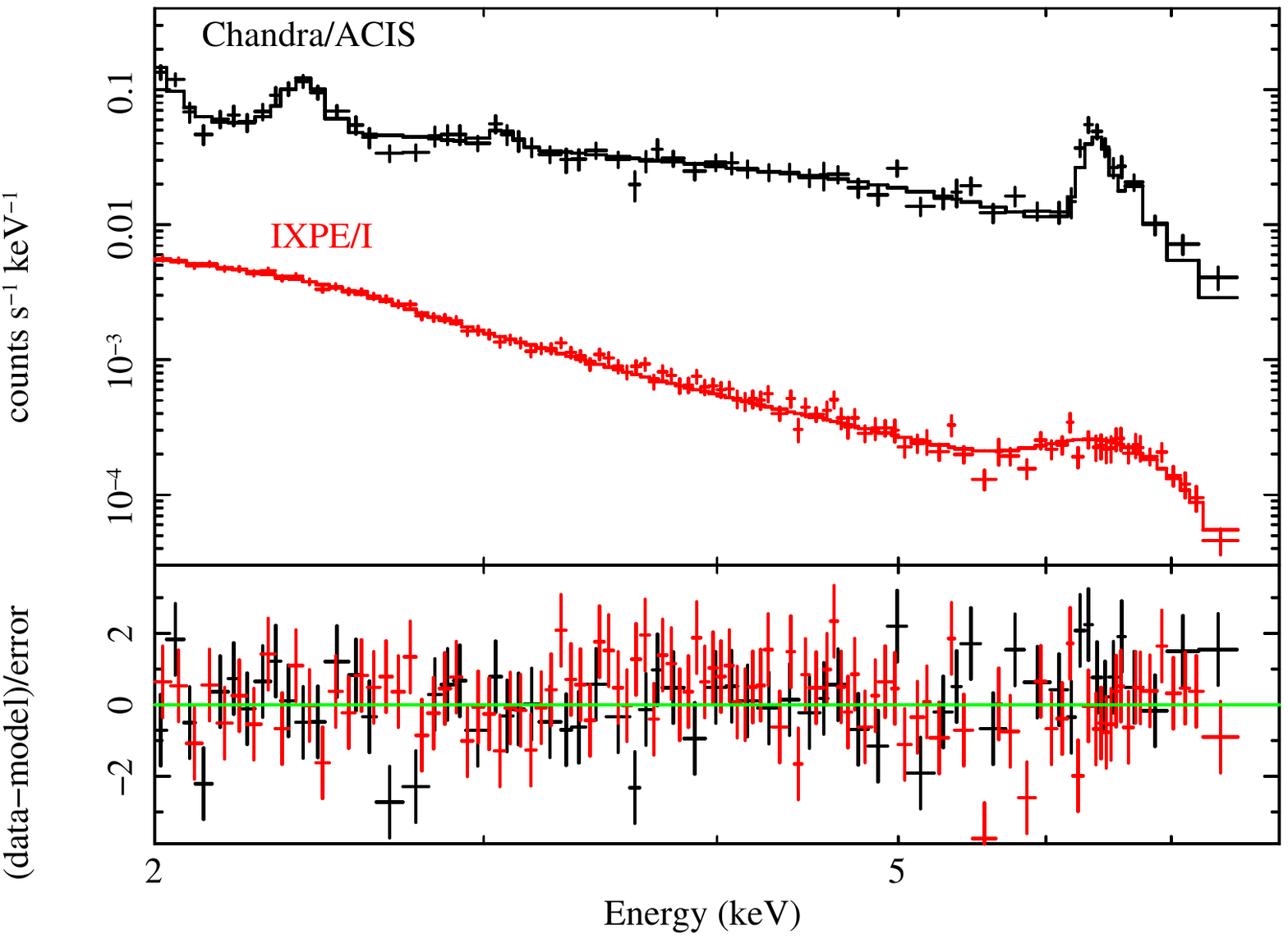}
\includegraphics[width=\columnwidth]{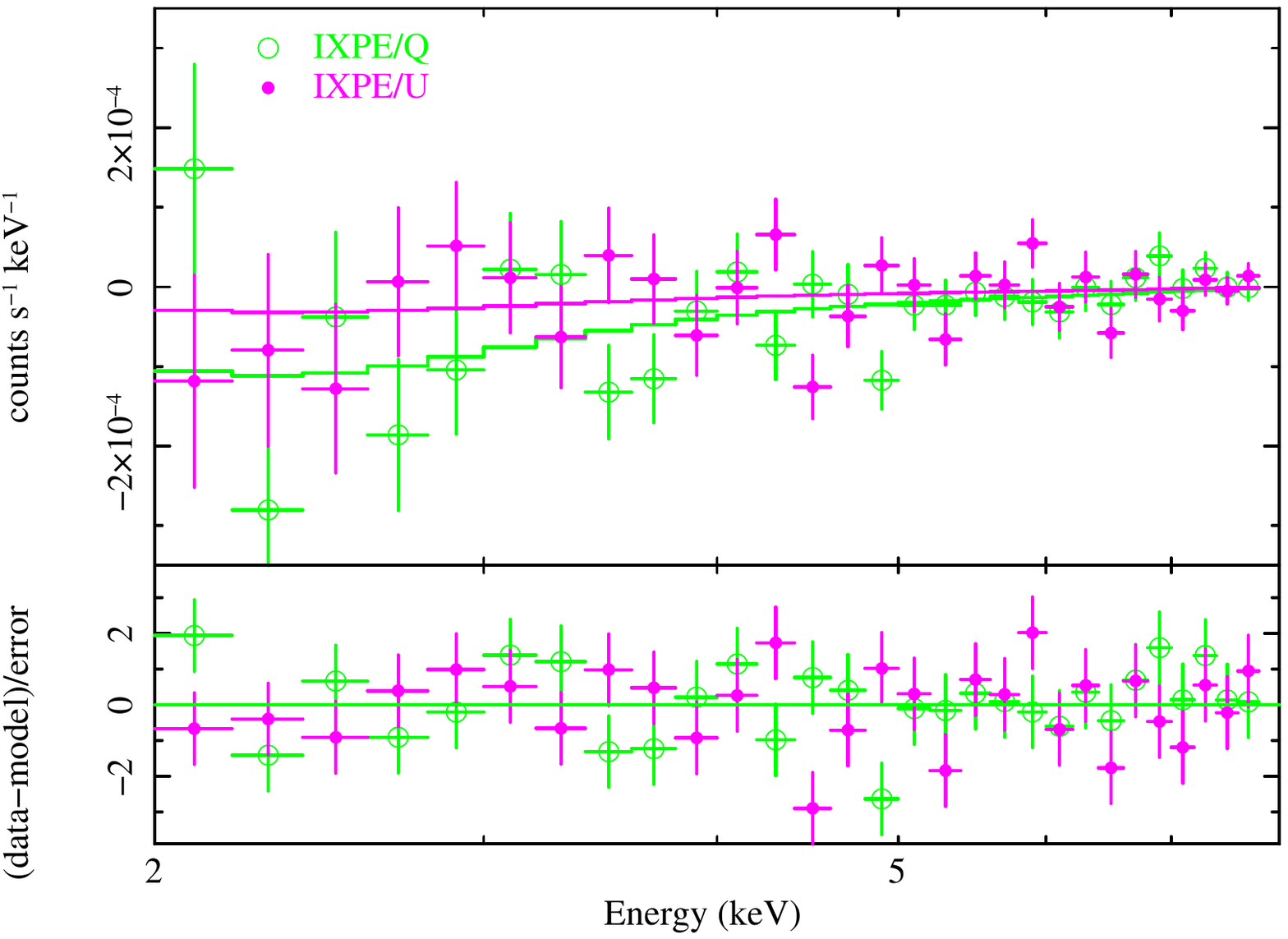}
\caption{
Top panel: \textit{Chandra}/ACIS and \textit{IXPE} $I$ spectra with best-fitting model and residuals.
Bottom panel: \textit{IXPE} $Q$ and $U$ Stokes spectra with best-fitting model and residuals. 
}
\label{fig:chandra+ixpe_data}
\end{figure}

\begin{figure}[]
\centering
\includegraphics[width=\columnwidth]{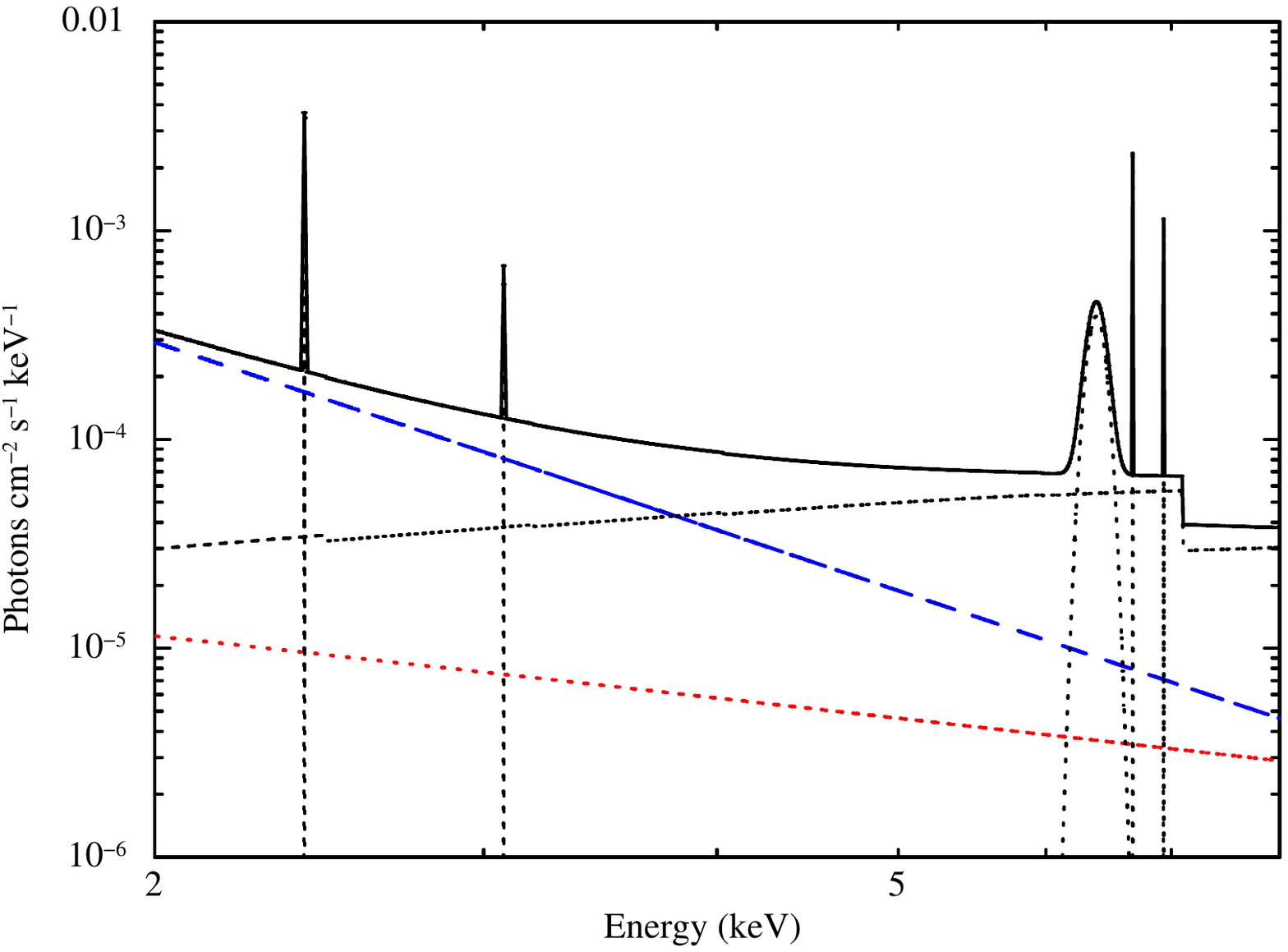}
\caption{
Best-fitting total model (black solid line) for the \textit{Chandra+IXPE} data. The plot shows the different components, i.e. cold reflection (black dotted line), warm reflection (blue dashed line), ULXs (red dotted line).
}
\label{fig:spectral_model}
\end{figure}

To fit the \textit{IXPE} spectra, we also include the power law describing the ULX emission as derived from \textit{Chandra}, fixing both the photon index and the flux. Finally, we multiply the different components by the {\sc polconst} model. We assume the emission lines and the ULXs to be unpolarized\footnote{ULXs are likely polarized, with $P$ depending on their true nature, orientation and geometry \citep{Veledina2023}. However, the ULXs around NGC~1068 contribute so weakly to the total flux that assuming no polarization is a reasonable hypothesis.}.
The {\sc xspec} model is as follows:
\begin{align*}
\textsc{c\_cal} \times
\textsc{tbabs} \times &[  
\textsc{polconst}^{(0)} 
\times 
(
\textstyle \sum \textsc{zgauss}^{(i)} && \text{lines} \\
&+ \textsc{powerlaw}) && \text{ULXs} \\
&+ \textsc{polconst}^{(w)} 
\times 
\textsc{powerlaw} && \text{warm refl.}\\
& + \textsc{polconst}^{(c)}
\times
\textsc{pexrav} 
] && \text{cold refl.} 
\end{align*}
where \textsc{c\_cal} is the cross-calibration constant. We fix the polarization degree of \textsc{polconst}$^{(0)}$ at zero, which must be set in \textsc{xspec} to describe an unpolarized component. The polarization angle of \textsc{polconst}$^{(0)}$ is also formally set at zero, even though this parameter has no meaning when the polarization degree is zero.\footnote{If we leave \textsc{polconst}$^{(0)}$ free to vary, we only obtain loose constraints, with a 1-sigma upper limit to the polarization degree of 30\%. We also tried to assess the contribution to the observed polarization by the emission lines and ULXs separately, via the multiplication of \textsc{polconst} with the \textsc{zgauss} and \textsc{powerlaw} components. In this case,  we obtain a 1-sigma upper limit of 60\% to the polarization of the ULXs component and of 40\% to the polarization of the lines.
}
The data and best-fitting model are shown in Figs.~\ref{fig:chandra+ixpe_data} and \ref{fig:spectral_model}. 

We find significant degeneracy among the polarization parameters of the warm and cold reflectors. The contour plot (not shown here) between the polarization degrees of the two components indicates that even at 1-sigma, $P_{\rm cold}$ is barely above zero and $P_{\rm warm}$ has only an upper limit of 28\%. To reduce such degeneracy, we need to make some assumptions. We thus fix the polarization parameters of the warm component using the far-UV polarimetric measurements by \cite{Antonucci1994}, see also Sect.~\ref{exploitation:comparison}. Those authors have shown that, within the first arcseconds, electron scattering in the winds is mainly responsible for the far-UV polarization. If electron scattering prevails in the wind, its scattering-induced polarization should be similar from the optical to the X-rays, so we postulate at first order that the X-ray counterpart could be similar. We thus fix the polarization degree and angle of the warm reflector at 16\% and 97$^\circ$, respectively. With this assumption, we obtain a good fit with $\chi^2 / \mathrm{dof}= 534/527$ : the cold reflector has a polarization degree of 20\% $\pm$ 10\% (non-zero at merely two sigma) and a polarization angle of 102$^\circ$ $\pm$ 15$^\circ$. The best-fitting paramters are reported in Table \ref{tab:chandra+ixpe_fit}.
Finally, we note that the warm reflection component is expected to be still significantly contaminated by unresolved emission lines, in which case the polarization degree could be much lower than our initial assumption. In the extreme hypothesis of null polarization of the warm reflector, we obtain a statistically equivalent fit, and a polarization degree of the cold reflector of $36\% \pm 10\%$ with a polarization angle similar to that obtained in the first case. 

\begin{table}
\label{table:1}      %
\centering                                     
\begin{tabular}{c c c c }          
\hline\hline                      
Energy (keV) & F (erg cm$^{-2}$ s$^{-1}$) & $P$ (\%) & $\Psi$ ($^\circ$) \\    
\hline        
2.0 -- 8.0 & $4.32\times10^{-12}$ & 12.4 $\pm$ 3.6 & 100.7 $\pm$ 8.3 \\     
\hline 
2.0 -- 3.5 & $1.03\times10^{-12}$ & 9.8 $\pm$ 4.9 & 104.2 $\pm$ 14.9 \\
3.5 -- 6.0 & $1.21\times10^{-12}$ & 21.3 $\pm$ 6.7 & 102.2 $\pm$ 9.2 \\
6.0 -- 8.0 & $2.08\times10^{-12}$ & $< 38.5$ & -- \\
\hline 
2.0 -- 5.5 & $1.97\times10^{-12}$ & 14.6 $\pm$ 3.9 & 101.1 $\pm$ 7.6 \\   
5.5 -- 8.0 & $2.35\times10^{-12}$ & $<26.8$ & -- \\
\hline                                             
\end{tabular}
\caption{{\it Chandra} fluxes F and {\sc xspec} measured polarization degree $P$ and angle $\Psi$ for various energy bands (obtained from the IXPE fit alone). Errors are given at 68\% confidence level for one parameter of interest. Upper limits are given at 99\% confidence level.} \label{tab:xspec_intvls}
\end{table}

\begin{table}
\begin{center}
  \begin{tabular}{ll}
\hline
\hline
Parameter & Value \\ 
\hline 
   \multicolumn{2}{c}{Lines (\textsc{zgauss})} \\
   $E_1$ (keV) & $1.96 \pm 0.02 $ \\
   $N_1$ & $5.6 \pm 0.8 \times 10^{-5}$ \\
   $E_2$ (keV) & $2.42 \pm 0.07$ \\
   $N_2$ & $3.5 \pm 0.3 \times 10^{-5}$ \\
   $E_3$ (keV) & $3.09^{+0.02}_{-0.04}$ \\
   $N_3$ & $0.55^{+0.19}_{-0.17} \times 10^{-5}$ \\
   $E_4$ (keV) & $6.41^{+0.01}_{-0.02}$ \\
   $\sigma_4$ (eV) & $76 \pm 12$ \\
   $N_4$ & $7.4^{+0.5}_{-0.6} \times 10^{-5}$ \\    
   $E_5$ (keV) & $6.7$(f) \\
   $N_5$ & $2.3 \pm 0.4 \times 10^{-5}$ \\
   $E_6$ (keV) & $6.966$(f) \\
   $N_6$ & $1.1 \pm 0.5 \times 10^{-5}$ \\   
   \multicolumn{2}{c}{ULXs (\textsc{powerlaw})} \\
   $\Gamma$ & $1$(f) \\
    $N$ & $2.3 \times 10^{-5}$(f) \\
   \multicolumn{2}{c}{Cold reflector (\textsc{pexrav})} \\
$\Gamma$ & $2$(f) \\
$N$ & $(1.97 \pm 0.08) \times 10^{-2}$ \\
P.D. (\%) & $20 \pm 10$ \\
P.A. (deg) & $102 \pm 15$\\
 \multicolumn{2}{c}{Warm reflector (\textsc{powerlaw})} \\
 $\Gamma$ & $3$(f) \\
 $N$ & $2.36^{+0.06}_{-0.12} \times 10^{-3}$ \\
P.D. (\%) & $16$(f)\\ 
P.A. (deg) & $97$(f) \\
\multicolumn{2}{c}{Cross-calibration constants} \\
$C_{\rm DU1-ACIS}$ & $0.68 \pm 0.01$ \\
$C_{\rm DU2-ACIS}$ & $0.64 \pm 0.01$  \\
$C_{\rm DU3-ACIS}$ & $0.60 \pm 0.01$ \\
\multicolumn{2}{c}{Observed flux} \\
$F_{\rm 2-8keV}$ & $(4.32 \pm 0.15) \times 10^{-12}$ \\
\hline
$\chi^2$/d.o.f. & $534/527$\\
\hline 
\end{tabular}
\end{center}
\caption{Best-fitting parameters (68 per cent confidence level for one parameter of interest) of the joint \textit{Chandra} and \textit{IXPE} fit. Normalizations are in units of photons keV$^{-1}$ cm$^{-2}$ s$^{-1}$, while the flux is in unit of erg cm$^{-2}$ s$^{-1}$. (f) denotes a fixed parameter. The energies of the two lines $E_5$ and $E_6$ are fixed at the values of the Fe~XXV and Fe~XXVI K$\alpha$ lines. The width of all lines is fixed at zero, with the only exception of the neutral Fe K$\alpha$ line.}
\label{tab:chandra+ixpe_fit}
\end{table}

\section{Analysis}
\label{exploitation}

\subsection{The geometrical distribution of scatterers}
\label{exploitation:Geometry}

The observed X-ray polarization angle can be compared to the position angle PA of the parsec-scale radio-jet seen in NGC~1068 to determine whether scattering occurs along the equatorial plane ($\Psi$ - PA $\approx$ 0$^\circ$) or in the polar direction ($\Psi$ - PA $\approx$ 90$^\circ$). The resolved parsec-scale jet on the 4.9~GHz map of \citet{Wilson1983} sustains a PA $\sim$ 34$^\circ$, but the jet does not follow a straight trajectory from the supposed position of the supermassive black hole (S1 component) to the terminal lobes. It is, in fact, deflected to the north east (see \citealt{Gallimore2004}). The central, sub-arcsecond jet structure was resolved by VLBA + phased-VLA 1.4~GHz observations and the inner PA of the jet (going from component S1 to C) is at about 12$^\circ$ (see Fig.~2 in \citealt{Gallimore2004} and also the very precise cartography of the radio emission in NGC~1068 by \citealt{Mutie2024}). The subtraction gives us 101$^\circ$ - 12$^\circ$ = 89$^\circ$. We thus conclude that the observed X-ray polarization angle of NGC~1068 is perpendicular to the radio structure axis\footnote{We note that the higher resolution VLBA images in \citet{Gallimore2004} show a torus structure for S1 (see their Figs.~3c and 4c), with PA 104.5$^\circ$ -- 108.1$^\circ$, again a very good match to the X-ray polarization angle we measured.}, implying that the observed X-ray polarization arises from scattering onto material that is preferentially situated well above the equatorial plane. 

Two components of the AGN can be responsible for this: the outflows or the torus. Our measurements reported in Tab.~\ref{tab:xspec_intvls} show that $\Psi$ is similar in the 2 -- 3.5~keV band (where the warm component dominates the X-ray spectrum, see Sect.~\ref{Results:IXPE+Chandra}) and in the 3.5 -- 6~KeV band, where the cold component takes precedence. Both components are thus likely sharing the polarization position angle. While producing perpendicular polarization in a polar wind is trivial \citep{Goosmann2007}, obtaining a similar polarization angle from an equatorial torus implies that scattering must occur inside the funnel and/or on the uppermost (outer) edges of this Compton-thick region. Such finding is strongly supported by the Atacama Large Millimeter Array ({\it ALMA}) observation of an extended patch of linear polarization arising from a spatially resolved elongated nuclear disk of dust of $\sim$ 50 -- 60~pc in diameter and oriented along an averaged PA of $\sim$ 113$^\circ$ \citep{Garcia2019}. 

The {\it Chandra} map at 0.5" resolution of the X-ray emission in NGC~1068 presented by \citet{Ogle2003} also supports our conclusion that the X-ray polarization we measured with {\it IXPE} mainly originates from the torus. The authors have shown (see their Figs.~3 and 4) that the peak of the 6 -- 8~keV (Fe~K) emission is coincident with the nucleus, as expected if produced in the inner wall of the molecular torus. Surrounding the X-ray peak in the nuclear region, 3 -- 6~keV extended emission is detected, which is attributed to X-rays that have scattered on the outer edge of the torus. The emission from the warm reflector (the ionization cones) is seen in their {\it Chandra} 1.3 -- 3~keV map and is dominated by emission from highly ionized Mg, Si, and S. The warm component is also appearing in their 3 -- 6~keV emission map but it is mostly scattered continuum. Using {\sc xspec} to measure the associated X-ray polarization in those three energy bands (see Tab.~\ref{tab:xspec_intvls}) gives only an upper limit to the 6 -- 8~keV emission, resulting from the combination of higher background levels and strong, little-to-no polarized emission from the iron K complex. In the 3.5 -- 6~keV band, where scattering off the torus is prevalent, $P$ rises up to $\sim$ 21\%. In the softest X-ray band (2 -- 3.5~keV), the observed polarization degree is much lower ($\sim$ 10\%), mostly due to the forest of narrow emission lines that depolarise the signal. Many of those lines are significantly enhanced by photo-excitation, so they might be polarized, albeit at a lower level from the hard continuum.

\subsection{Estimating the AGN inclination and torus geometry}
\label{exploitation:morphology}

 Even if nothing definitive can be said about the intrinsic polarization degree and angle originating from the torus, we can try to evaluate the geometry of this region thanks to Monte Carlo radiative transfer simulation of optically thick tori. We used the radiative transfer Monte Carlo code {\sc stokes} \citep{Goosmann2007,Marin2012,Marin2015,Marin2018b,Rojas2018} to simulate a central, isotropical, continuum source with a power law index $\Gamma$ $\sim$ 2.04 (\citealt{Matt2004},\citealt{Pounds2006} and Sect.~\ref{Results:IXPE+Chandra}). The source emits a 2\% parallel polarized primary continuum, since it has been shown in NGC~4151 that the X-ray continuum is parallelly polarized by a few percents \citep{Gianolli2023}. Around the central source, we modeled a uniform-density torus using a circular cross-section. The inner wall of the torus is set at a fixed distance of r$_{in}$ = 0.25~pc \citep{Lopez2018,Vermot2021} and the radius $a$ of the torus is set by the region's half-opening angle $\Theta$ such as $a$ = r$_{in}$ $\cos$ $\Theta$/(1-$\cos$ $\Theta$). The neutral hydrogen density N$_{\rm H} $ is set to 10$^{25}$~atom~cm$^{-2}$ along the equatorial plane \citep{Matt2004}. The half-opening angle of the torus and the system inclination $i$ are free parameters, both measured from the vertical symmetry axis of the system. Type-2 AGNs are thus found for $i$ $>$ $\Theta$. Such models have been examined in details in \citet{Podgorny2023} and we refer the reader to this publication for in-depth details.

\begin{figure}
    \centering
    \includegraphics[trim= 0 0 1cm 1cm, clip, width=.5\textwidth]{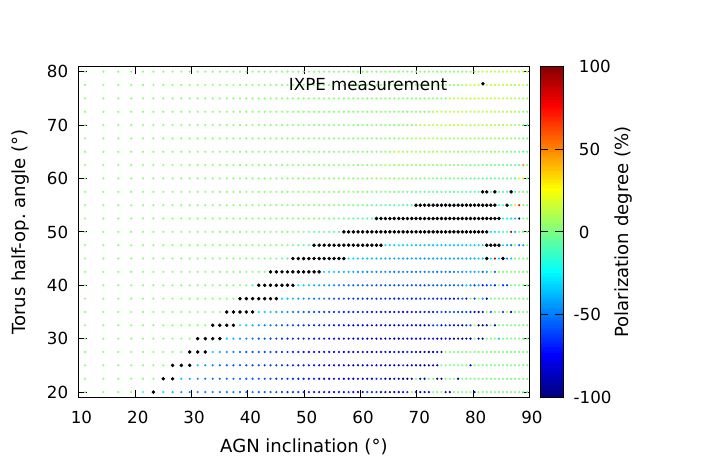} 
    \caption{{\sc stokes} simulations of a circularly-shaped torus with various half-opening angles $\Theta$ seen at several observer inclinations $i$. The color-coded polarization degree is integrated from 3.5 to 6.0~keV to avoid contribution from intense emission lines. A positive polarization degree indicates that the photon electric field vector is parallel to the axis of symmetry of the system. Thus, negative polarization stands for perpendicularly polarized photons. Each colored dot represents a ($\Theta$,$i$) simulation. If the polarization degree and angle of a dot is consistent with the X-ray polarization of the torus as measured by {\it IXPE}, the dot is shown in black.}
    \label{Fig:Simulations}
\end{figure}

We present in Fig.~\ref{Fig:Simulations} the results of the simulations for a wide variety of $\Theta$ and $i$ angles. The polarization is integrated between 3.5 and 6~keV to avoid as much contamination as possible from the fluorescent and recombination lines (see Fig.~\ref{fig:spectral_model}). Superimposing the estimated neutral reflector X-ray polarization (20\% $\pm$ 10\%) to the models allows us to put strong constraints on the geometrical configuration of the system. Excluding a) all torus half-opening angles smaller than that of the ionized polar winds seen in NGC~1068, that sustain a half-opening angle of 40$^\circ$ as traced by [O III] emission \citep{Macchetto1994}, b) all results for $i$ $<$ $\Theta$ and c) all non-perpendicular polarization angles (an assumption compatible with {\it IXPE} data), we find that the range of permitted AGN inclinations is 42$^\circ$ -- 87$^\circ$ and the range of torus half-opening angles is 40$^\circ$ -- 57.5$^\circ$ (as illustrated by the black dots on Fig.~\ref{Fig:Simulations}). Further observation with better statistics could narrow down the parameter space but, if we rely on the range of inclinations $i$ estimated from the literature (70$^\circ$ -- 90$^\circ$, see Sect.~\ref{Introduction}), the permitted range of $\Theta$ becomes 45$^\circ$ -- 57.5$^\circ$, with a much higher probability to be in the range 50$^\circ$ -- 55$^\circ$. Interestingly, this is comparable to the range of $\Theta$ found in the case of the Circinus galaxy (45$^\circ$ - 55$^\circ$), the only other type-2 AGN probed by {\it IXPE} \citep{Ursini2023}.

For the sake of completeness, we also tested the line dominated warm reflector scenario ($P$ = 0\%), that would lead to a neutral reflector polarization of $36\% \pm 10\%$ (see Sect.~\ref{Results:IXPE+Chandra}). By doing so, the torus half-opening angle estimated from Monte Carlo radiative transfer simulations becomes 40$^\circ$ -- 50$^\circ$, with a much higher statistical probability to be between 45$^\circ$ and 50$^\circ$. This is also comparable to the range of $\Theta$ found in the case of the Circinus galaxy.

\subsection{Comparison with optical data}
\label{exploitation:comparison}

The broadband polarization of NGC~1068 has been thoroughly examined from the ultraviolet to the infrared band (see \citealt{Marin2018} and reference therein for a review). Yet, due to strong starlight dilution by the host, the true (intrinsic) polarization levels of the AGN is difficult to estimate. It is possible to correct the UV-optical polarization for dilution by subtracting out the stellar component(s) from the total flux, but it is an uneasy process. The addition of the X-ray polarization measurement achieved by {\it IXPE} brings another piece to the puzzle, since it is essentially free from diluting sources (albeit the almost negligible ULX signal). A comparison between X and optical data can thus lead to better constraints on the global geometry of the AGN. 

To do so, 0.35 -- 1.1~$\mu$m linear spectropolarimetry was obtained on December 1 -- 6th, 2023, with the FOcal Reducer/low dispersion Spectrograph 2 (FORS2) instrument mounted on the Very Large Telescope (VLT), quasi-simultaneously to the {\it IXPE} observation. The observation, under program ID 112.26WL.001, will be analyzed and published later on (Marin et al., in prep.), but the reduced optical data confirm that the linear polarization from NGC~1068 remained stable since the last decades, especially in the continuum. The polarization degree rises from $\sim$ 1.5\% in the red up to $\sim$ 12\% in the blue band. The polarization position angle is almost constant with wavelength, slowly rotating from $\sim$ 100$^\circ$ in the red to $\sim$ 90$^\circ$ in the blue. The rise of $P$ continues in the ultraviolet band up to 16\% at a position angle of 97$^\circ$ shortward of 2700~\AA~\citep{Code1993,Antonucci1994}. The polarization degree then plateau, as the host starlight contribution becomes negligible, indicating that electron scattering dominates in the winds.

The matching of the polarization angle between the near-infrared, optical, ultraviolet and X-rays clearly indicates that the polarization results, in all cases, from scattering by material well above the equatorial plane. But they are not due to the same AGN component. X-ray spectropolarimetry tends to favor light reflection off atoms and molecules from the cold component (the torus), while optical imaging and spectropolarimetry advocates for electron (or dust) scattering inside the polar outflows \citep{Antonucci1985,Antonucci1994}. In the spectral region where emission from the ionized winds dominates the X-ray signal (2 -- 3.5~keV, see Fig.~\ref{fig:spectral_model}), we measured $P$ = 9.8\% $\pm$ 4.9\% at 104.2$^\circ$ $\pm$ 14.9$^\circ$. Although diluted by the emission lines, the polarized X-ray continuum probably comes from the optically thin region that scatters the optical and ultraviolet photons \citep{Antonucci1985}. 

As things currently stand and with the statistics offered by our observation, we note that the optical/ultraviolet and X-ray measurements are different in $P$ (but not in $\Psi$), so we caution using ultraviolet polarization measurements of Seyfert-2 galaxies as a basic proxy to predict the polarization level of a source in the X-rays.

\section{Conclusions}
\label{Conclusions}
We have measured the 2 -- 8~keV polarization of the archetypal Seyfert-2 AGN NGC~1068 with {\it IXPE}. We found a linear polarization degree of 12.4\% $\pm$ 3.6\% at 100.7$^\circ$ $\pm$ 8.3$^\circ$ (68\% confidence level). The polarization is perpendicular to the position angle of the parsec-scale radio structure, similarly to what is observed for all Seyfert-2s in the optical band. A combined {\it Chandra} and {\it IXPE} analysis indicated a significant degeneracy among the polarization parameters of the warm and cold reflectors but, from multi-wavelength constraints, we estimated that the cold reflector has a polarization degree of 20\% $\pm$ 10\% and polarization angle of 102$^\circ$ $\pm$ 15$^\circ$. By taking the result at the face value, numerical simulations allowed us to derive a probable torus half-opening angle of 50 -- 55$^\circ$ (from the vertical axis of the system). This morphological constraint is quite similar to the torus half-opening angle derived for the Circinus galaxy thanks to X-ray polarimetry, the only other type-2 AGN probed by {\it IXPE} so far. By doing so, X-ray polarimetry has shown its powerful and unique capabilities of constraining the geometrical arrangement of matter around supermassive black holes.

~\

\textbf{Acknowledgments}. The Imaging X-ray Polarimetry Explorer ({\it IXPE}) is a joint US and Italian mission. The US contribution is supported by the National Aeronautics and Space Administration (NASA) and led and managed by its Marshall Space Flight Center (MSFC), with industry partner Ball Aerospace (contract NNM15AA18C).  The Italian contribution is supported by the Italian Space Agency (Agenzia Spaziale Italiana, ASI) through contract ASI-OHBI-2022-13-I.0, agreements ASI-INAF-2022-19-HH.0 and ASI-INFN-2017.13-H0, and its Space Science Data Center (SSDC) with agreements ASI-INAF-2022-14-HH.0 and ASI-INFN 2021-43-HH.0, and by the Istituto Nazionale di Astrofisica (INAF) and the Istituto Nazionale di Fisica Nucleare (INFN) in Italy.  This research used data products provided by the {\it IXPE} Team (MSFC, SSDC, INAF, and INFN) and distributed with additional software tools by the High-Energy Astrophysics Science Archive Research Center (HEASARC), at NASA Goddard Space Flight Center (GSFC). F.M. thanks Robert Antonnucci, Makoto Kishimoto, Patrick Ogle and Ari Laor for their valuable comments on the manuscript. A.D.M., E.Co., R.F., S.F., F.L.M., F. Mu. P.So. are partially supported by MAECI with grant CN24GR08 “GRBAXP: Guangxi-Rome Bilateral Agreement for X-ray Polarimetry in Astrophysics”. F.T. and M.L. acknowledge funding from the European Union - Next Generation EU, PRIN/MUR 2022 (2022K9N5B4). M.D., J.P., and V.K. thank GACR project 21-06825X for the support and institutional support from RVO:67985815. I.L. was supported by the NASA Postdoctoral Program at the Marshall Space Flight Center, administered by Oak Ridge Associated Universities under contract with NASA. The work of RTa and RTu is partially supported by grant PRIN-2022LWPEXW of the Italian MUR. The French contribution (F.M., T.B., V.E.G., P.-O.P.) is partly supported by the French Space Agency (Centre National d’Etude Spatiale, CNES) and by the High Energy National Programme (PNHE) of the Centre National de la Recherche Scientifique (CNRS). The authors thank the anonymous referee for her/his helpful comments that improved the quality of the manuscript.


\appendix

\section{Summary of PCUBE analysis}
\label{Annex}

In order not to confuse the reader with both the PCUBE and \textsc{XSPEC} results in the main body of the article, we summarize here our findings using a PCUBE analysis. We computed the significance of detection for a variety of energy bins, calculated in the same way presented in Sect.~\ref{Results:IXPE_pcubes}, considering two degrees of freedom and following the official IXPE statistical guide (\url{https://heasarc.gsfc.nasa.gov/docs/ixpe/analysis/IXPE_Stats-Advice.pdf}). We also detail the variability observed in both time and energy domains for three separate bins. Our results can be found in Tab.~\ref{Tab:appendix}.

\begin{table*}
\label{Tab:appendix}      %
\centering                                     
\begin{tabular}{c c c c }          
\hline\hline                      
Energy (keV) & Detection Significance (\%) & $P$ (\%) & $\Psi$ ($^\circ$) \\    
\hline        
2.0 -- 3.5 & 90.6 & 10.6 $\pm$ 4.9 & 111.2 $\pm$ 13.2 \\    
3.5 -- 6.0 & 98.3 & 20.0 $\pm$ 7.0 & 94.8 $\pm$ 10.0 \\  
6.0 -- 8.0 & 90.6 & 13.6 $\pm$ 12.5 & 156.5 $\pm$ 26.3 \\  
\hline 
2.0 -- 5.5 & 99.7 & 14.1 $\pm$ 4.1 & 105.1 $\pm$ 8.4 \\
5.5 -- 8.0 & 26.1 & 8.7 $\pm$ 11.1 & 156.9 $\pm$ 36.8 \\
\hline 
2.0 -- 4.0 & 96.2 & 11.4 $\pm$ 4.5 & 107.1 $\pm$ 11.2 \\   
4.0 -- 6.0 & 93.9 & 19.9 $\pm$ 8.4 & 95.8 $\pm$ 12.1 \\  
\hline 
2.0 -- 6.0 & 99.5 & 13.5 $\pm$ 4.2 & 102.5 $\pm$ 8.9 \\  
\hline 
2.0 -- 8.0 & 82.1 & 9.5 $\pm$ 5.1 & 112.9 $\pm$ 15.5 \\ 
\hline                                             
\end{tabular}
\caption{Results from our PCUBE analysis. The first column is the energy binning (in keV), the second column the detection significance, the third column is the  polarization degree (with one-dimensional errors) and the fourth column is the polarization angle.} 
\end{table*}

As shown in the table, the most solid detections are found in the 2 -- 5.5 and 2 -- 6~keV bands, before the onset of fluorescent emission from the iron K complex and its associated dilution. The values measured in those energy bands are above the minimum detectable polarization, that is the degree of polarization corresponding to the amplitude of
modulation that has only a 1\% probability of being detected by chance \citep{Weisskopf2010}.

\begin{figure}
    \centering
    \includegraphics[width=.5\textwidth]{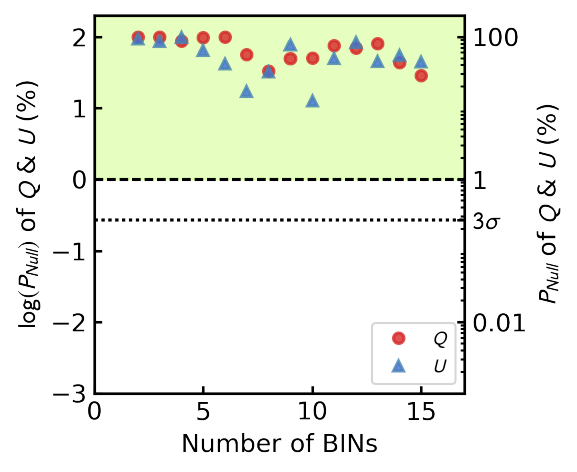}
    \quad
    \includegraphics[width=.5\textwidth]
    {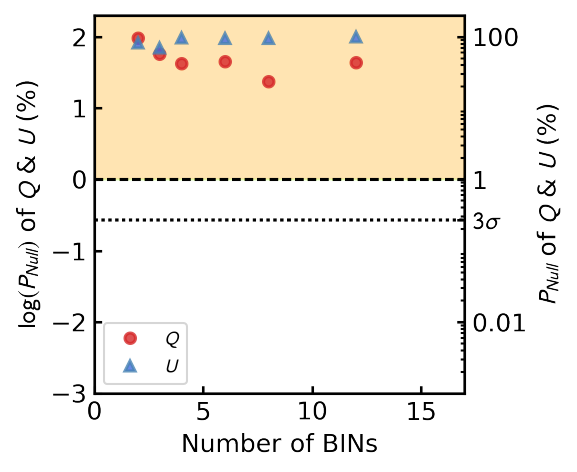}
    \caption{The null hypothesis probability of the chi-square test for the constant model on the normalized $Q$ (red) and $U$ (blue) Stokes parameters for different time (top panel) and energy (bottom panel) binning cases. The left and right vertical axes in each panel show the percentage of the probability values on a logarithmic and linear scale, respectively. The green and orange shaded areas indicate the 1\% threshold level. The black dashed and dotted lines in the middle of each panel represent the 1\% and 3$\sigma$ (99.73\%) probability levels, respectively.}
    \label{Fig:EnergyBins}
\end{figure}

Regarding the variability tests, mentioned in Sect.~\ref{Results:IXPE_pcubes}, we show in Fig.~\ref{Fig:EnergyBins} the null hypothesis tests of polarization variability over time (top) and energy (bottom). The lowest probability for the time binning case is 27.63\%, and for the energy binning case is 23.80\%. In all cases, the polarization appears consistent with being constant in time and energy.

\end{document}